\documentclass[%
 reprint,
 amsmath,amssymb,
 aps,
]{revtex4-2}

\usepackage{graphicx}
\usepackage{dcolumn}
\usepackage{bm}
\usepackage[scr=rsfs]{mathalpha}
\usepackage{mathtools} 
\usepackage{booktabs}
\usepackage{hyperref}
\usepackage{url}
\begin{document}

\preprint{APS/123-QED}

\title{Reconstruction of Stochastic Dynamics from Large Streamed Datasets}

\author{William Davis}
 \email{w1davis@ucsd.edu}
 \affiliation{Cecil H. and Ida M. Green Institute of Geophysics and Planetary Physics, Scripps Institution of Oceanography, University of California, San Diego, La Jolla, California 92037, USA}

\date{\today}

\begin{abstract}
The complex dynamics of physical systems can often be modeled with stochastic differential equations. 
However, computational constraints inhibit the estimation of dynamics from large time-series datasets.
I present a method for estimating drift and diffusion functions from inordinately large datasets through the use of incremental, online, updating statistics. 
I demonstrate the validity and utility of this method by analyzing three large, varied synthetic datasets, as well as an empirical turbulence dataset. 
This method will hopefully facilitate the analysis of complex systems from exceedingly large, ``big data'' scientific datasets, as well as real-time streamed data.

\end{abstract}

\maketitle


\section{\label{sec:intro}Introduction}

The dynamics of complex systems with many degrees of freedom can often be modeled as continuous-time stochastic processes~\cite{haken2004synergetics}.
When a system is modeled by a stochastically forced, scalar, first-order differential equation, the temporal evolution of a quantity~$X(t)$ is described by a Langevin-type equation~\cite{risken1996fokker}: 

\begin{equation}\label{eq:single-order-parameter-sde}
    \frac{d}{dt}X(t) = f(X) + g(X)\Gamma(t),
\end{equation}
where a separation of scales partitions the dynamics of $X(t)$ into slow changes modulated by $f(X)$, and rapidly-varying changes modulated by $g(X)$.
Fluctuations are driven by Gaussian white noise~$\Gamma(t)$, with~$\langle\Gamma(t)\rangle=0$ and~$\langle\Gamma(t)\Gamma(t^\prime)\rangle=\delta(t-t^\prime)$. 
Here and throughout, the It\^{o} interpretation is adopted.
If~$X(t)$ contains no discontinuous jumps, then the evolution of the probability density function can be described by the Fokker-Planck equation~\cite{risken1996fokker}

\begin{multline}\label{eq:fokker-planck-equation}
    \frac{\partial}{\partial t} p(x,t|x^\prime,t^\prime) = \Bigg[-\frac{\partial}{\partial x} D^{(1)}(x) \\
    + \frac{\partial^2}{\partial x^2} D^{(2)}(x) \Bigg] p(x,t|x^\prime,t^\prime)
\end{multline}
where~$p(\circ|\circ)$ is the transition probability, and~$x$ and~$x^\prime$ are state variables of~$X$. 
The Fokker-Planck equation contains the Kramers-Moyal (KM) coefficients

\protect 
\begin{equation}\label{eq:kramers-moyal-coefficients}
    D^{(k)}(x) = \lim_{\tau\rightarrow0} \frac{1}{k!\tau} 
    \int_{-\infty}^{\infty} \big[x^\prime - x\big]^k
    p(x^\prime,t+\tau|x,t)\ dx^\prime.
\end{equation}

The~$k=1$ and~$k=2$ KM coefficients are called the drift and diffusion functions, respectively, and they correspond to terms in the dynamical equation~\eqref{eq:single-order-parameter-sde}, with~$f(x)=D^{(1)}(x)$ and~$g(x)=\sqrt{2D^{(2)}(x)}$.

It has been shown that KM coefficients---and hence drift and diffusion functions---can be estimated from empirical samples of~$X(t)$, using a conditional averaging technique called ``direct estimation''~\cite{siegert1998analysis,gottschall2008definition}.
Direct estimation and descendant methods \cite[e.g.,][]{bottcher2006reconstruction,lind2010extracting,scholz2017parameter,lade2009finite,honisch2011estimation,rydin2021arbitrary,lehle2018analyzing} have been applied to time-series data in various fields of science~\cite{friedrich2011approaching,tabar2019analysis}, including 
turbulence~\cite{friedrich1997description,renner2001experimental,friedrich2020generalized},
wind energy~\cite{milan2013turbulent},
climate data~\cite{lind2005reducing,lind2010extracting}, and
geomagnetic field variations~\cite{buffett2013stochastic,davis2021inferring}.

Although the calculation of KM coefficients is conceptually simple~\cite{siegert1998analysis,lamouroux2009kernel}, estimations are prone to bias, especially in areas of rarely sampled state space~\cite{gorjao2019kramersmoyal}.
Inaccuracies are particularly apparent for processes with heavy tails, or for systems that exhibit rare, transient dynamics.
Attempts to resolve KM coefficients in rarely sampled regions by reducing the resolution of conditioning also result in biased drift and diffusion estimates~\cite{lamouroux2009kernel}.

A rudimentary but effective solution to the sampling problem is to perform analyses on datasets that are as large as possible. 
This approach is effective because the estimation bias of KM coefficients scales as~$1/\sqrt{N\Delta t}$, where~$N$ is the number of samples and~$\Delta t$ is the sampling interval~\cite{kleinhans2007quantitative}.
Indeed, in the era of ``big data,'' there is growing interest in estimating drift and diffusion functions for increasingly large scientific datasets~\cite{raischel2014human,tabar2019analysis}.
However, existing KM procedures calculate KM coefficients using offline methods~\cite{rinn2016langevin,gorjao2019kramersmoyal,gorjao2023jumpdiff}, requiring the complete dataset to be available at once, and with memory requirements that scale with the number of data points.
Large datasets are often incompatible with offline methods, either because the data cannot fit into computer memory, or because the data originates from arbitrarily large data streams~\cite{wang2016statistical}. 
An alternative approach is to use online methods, which incrementally update statistical estimates from streamed data, arriving one data point at a time~\cite{karp1992line,day2020onlinestats}.
In this paper I present an online method of computing KM coefficients from streamed time-series data, enabling the estimation of drift and diffusion functions from large time-series datasets which are unreachable with previous methods.

\section{\label{sec:moment-estimation}Estimation of Conditional Moments}

Consider a finite sample of~$N$ points in~$X(t)$ from process~\eqref{eq:single-order-parameter-sde}, denoted as 

\begin{equation}
    \mathscr{S}_N := \{ (t_1,X_1), (t_2,X_2), \dots, (t_N,X_N)\}.
\end{equation}
Here I assume a regular sampling interval~$\Delta t$.
The aim is to use these data to construct non-parametric estimates of drift and diffusion coefficients of the Langevin-type equation that generated $X(t)$.
Estimation of drift and diffusion coefficients is conducted at a set of~$N_x$ evaluation points in~$x$, represented by the vector

\begin{equation}\label{eq:x-vector}
    \bm{\mathcal{X}} := \left[x_1, x_2, \dots, x_{N_x}\right].
\end{equation}
Drift and diffusion estimates at these points will be denoted by the vector~$\mathbf{\hat{D}}^{(k)}$, with~$\hat{D}^{(k)}_j := \hat{D}^{(k)}(\mathcal{X}_j)$.
Estimation of $\mathbf{\hat{D}}^{(k)}$ requires evaluation of the conditional process increments---or ``conditional moments''~\cite{honisch2011estimation}---in~\eqref{eq:kramers-moyal-coefficients}, namely

\begin{equation}\label{eq:finite-time-CM}
    M^{(k)}(\tau,x) = \int_{-\infty}^\infty [x^\prime - x\big]^k p(x^\prime, t + \tau| x,t)\ dx^\prime,
\end{equation}
for~$k=1,2$.
As the~$\tau\rightarrow 0$ limit in~\eqref{eq:kramers-moyal-coefficients} cannot be performed for empirical data, \eqref{eq:finite-time-CM} is estimated at a set of~$N_\tau$ evaluation points in~$\tau$ values, represented by the vector

\begin{equation}\label{eq:tau-vector}
    \bm{\mathcal{T}} := \left[\Delta t, 2\Delta t, \dots, N_\tau\Delta t\right]^T.
\end{equation}
Estimates of conditional moments~\eqref{eq:finite-time-CM} are performed at all points in~$\bm{\mathcal{T}}$ and~$\bm{\mathcal{X}}$, and will be denoted as~${N_\tau \times N_x}$ matrices~$\mathbf{\hat{M}}^{(k)}$, with~$\hat{M}^{(k)}_{ij} := \hat{M}^{(k)}(\mathcal{T}_i,\mathcal{X}_j)$.
I now outline an existing estimation procedure for conditional moments, before proposing online updating formulas.

\subsection{Offline calculation}

One method of estimating conditional moments is kernel-based regression (KBR)~\cite{lamouroux2009kernel}.
A chosen kernel function~$K(\cdot)$ applies conditioning on the state variable,~$x$, and, assuming ergodicity, the estimators for \eqref{eq:finite-time-CM} can be written as

\begin{equation}\label{eq:KBR}
    \hat{M}^{(k)}_{ij} = \frac{
    \sum\limits_{n=1}^{N-i} K_h(\mathcal{X}_j - X_n)
    \big[X_{n+i} - X_n\big]^k}{
    \sum\limits_{n=1}^{N-i} K_h(\mathcal{X}_j - X_n)},
\end{equation}
for~$k=1,2$, where~$K_h(\cdot) = K(\cdot/h)/h$ is a scaling of the kernel with bandwidth~$h$.
The methods presented here are independent of the chosen kernel. Here I use the Epanechnikov kernel~\cite{epanechnikov1969non} 

\begin{equation}\label{eq:epan-kernel}
    K(x) = 
    \begin{cases}
        \frac34(1 - x^2) & \text{if~$x^2<1$}, \\
        0 & \text{otherwise},
    \end{cases}
\end{equation}
which has computationally favorable properties~\cite{hardle2004nonparametric}.
If kernel conditioning is replaced with bin counting, the estimation becomes histogram-based regression (HBR)~\cite{siegert1998analysis}.
Some studies also analyze the variance of the conditional process increments~(e.g.,~\cite[]{ragwitz2001indispensable,siefert2003quantitative,lehle2018analyzing}).
I will refer to this quantity as the ``conditional variance,'' and denote it as 
\begin{equation}\label{eq:KBR-variance}
    \hat{M}^{(2^*)}_{ij} = \frac{
    \sum\limits_{n=1}^{N-i} K_h(\mathcal{X}_j - X_n)
    \left(\big[X_{n+i} - X_n\big] - \hat{M}^{(1)}_{ij}\right)^2}{
    \sum\limits_{n=1}^{N-i} K_h(\mathcal{X}_j - X_n)}.
\end{equation}

Both HBR and KBR are implemented in modern software libraries~\cite{rinn2016langevin,gorjao2019kramersmoyal,fuchs2022open,gorjao2023jumpdiff}, and can be extended to irregularly sampled time-series data~\cite{davis2022estimation}.
However, these offline methods require the entire input~$\mathscr{S}_N$ to be available at once: the entire calculation must be repeated if more data is appended to~$\mathscr{S}_N$.

\subsection{Online calculation}

I now present formulas for updating sample conditional moments~\eqref{eq:KBR}, previously calculated from~$\mathscr{S}_{N-1}$, with a single new observation~$(t_N, X_N)$.
I refer to this approach as ``online kernel-based regression'' (OKBR).
To facilitate indexing, the subscript notation~$[\dots]\big|_n$ denotes a quantity calculated from the first~$n$ observations.
The updating formulas are written as (see Appendix)

\begin{multline}\label{eq:OKBR}
    \hat{M}^{(k)}_{ij}\big|_{N} = \hat{M}^{(k)}_{ij}\big|_{N-1} + K_h(\mathcal{X}_j - X_{N-i})\\
    \times\left(\left[X_N - X_{N-i}\right]^k - \hat{M}^{(k)}_{ij}\big|_{N-1}\right)\Big/W_{ij}\big|_{N},
\end{multline}
for~$k=1,2$, where~$W_{ij}\big|_{N}$ are cumulative weights 

\begin{equation}\label{eq:OKBR-W}
    W_{ij}\big|_{N} = W_{ij}\big|_{N-1} + K_h\left(\mathcal{X}_j - X_{N-i}\right).
\end{equation}
To define a corresponding updating formula for~\eqref{eq:KBR-variance}, I introduce the intermediate quantity~$S_{ij}\big|_{N}$, which corresponds to the weighted sum of squares of differences from the current mean,

\begin{equation}\label{eq:S-definition-offline}
    S_{ij}\big|_{N} := \sum\limits_{n=1}^{N-i} K_h(\mathcal{X}_j - X_n)
            \Big(\big[X_{n+i} - X_{n}\big] - \hat{M}^{(1)}_{ij}\big|_{N}\Big)^2,
\end{equation}
and is related to~\eqref{eq:KBR-variance} by 

\begin{equation}\label{eq:Mstar-relation}
    \hat{M}^{(2^*)}_{ij}\big|_{N} = S_{ij}\big|_{N}\bigg/W_{ij}\big|_{N}.
\end{equation}
The corresponding online formula is (see Appendix~\ref{app:ssec:variance-derivation})

\begin{multline}\label{eq:OKBR-momentx-variance}
    S_{ij}\big|_{N} = S_{ij}\big|_{N-1} + K_h(\mathcal{X}_j - X_{N-i})\\
    \times\left(\left[X_N - X_{N-i}\right] - \hat{M}^{(1)}_{ij}\big|_{N-1}\right)\\
    \times\left(\left[X_N - X_{N-i}\right] - \hat{M}^{(1)}_{ij}\big|_{N}\right),
\end{multline}
These formulas have been constructed to avoid numerical instability and loss of precision~\cite{welford1962note,west1979updating}.
In the next section, I validate the presented methods on three synthetic datasets.

\section{\label{sec:numerical-examples}Numerical Examples}

\subsection{\label{ssec:OU-process-example}Ornstein-Uhlenbeck process}

\begin{figure}
\noindent
\includegraphics[width=0.45\textwidth]{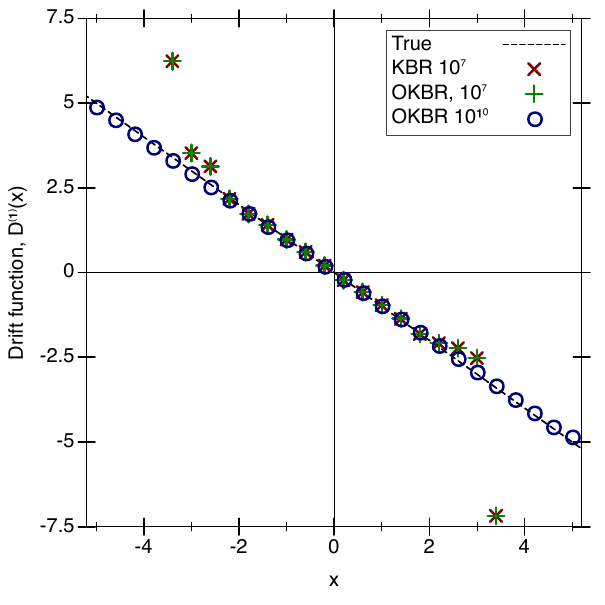}
\includegraphics[width=0.45\textwidth]{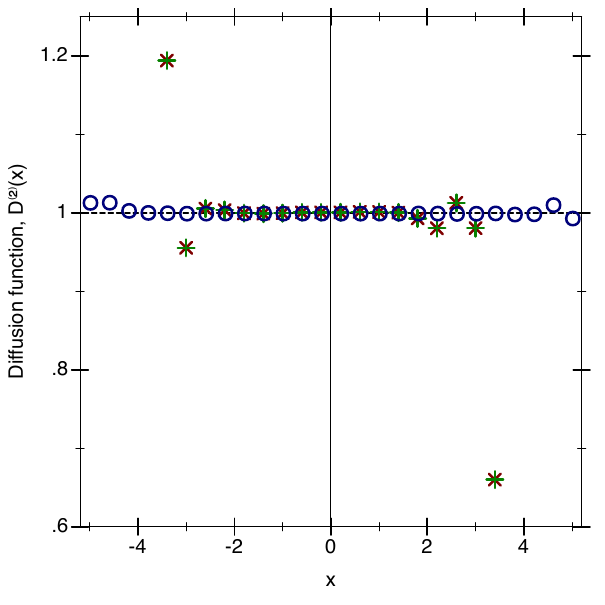}
\caption{Drift (upper) and diffusion (lower) estimates for example~\ref{ssec:OU-process-example}. 
The true drift and diffusion functions are indicated with dashed black lines.
Estimates from~$N=10^7$ data-points conducted with KBR and OKBR are shown with red ``$\times$''~crosses and green ``$+$''~crosses, respectively. 
Estimates from~$N=10^9$ data points conducted with OKBR are shown with open blue circles. 
For the~$N=10^7$ simulation, estimates of drift and diffusion coefficients at~$|x|\gtrsim 2$ are noisy or non-existent, due to sparse sampling. 
}
\label{figure:OU-example}
\end{figure}

I examine a simple example where the drift and diffusion functions are set as

\begin{subequations}\label{eq:OU-drift-noise}
\begin{align}
    D^{(1)}(x) &= -x,\\
    D^{(2)}(x) &= 1.
\end{align}
\end{subequations}
I numerically integrate~\cite{mil1975approximate} this process using a sampling interval of~$\Delta t=10^{-3}$ for~$N=10^7$ data-points.
I estimate conditional moments at 26 equally-spaced points in the range~$[-5,5]$ using a bandwidth of~$h=0.4$, and perform time sampling at a single time-step~$\bm{\mathcal{T}}=[\Delta t]$.
I conduct estimation using both the KBR formulas~\eqref{eq:KBR} and OKBR formulas~\eqref{eq:OKBR}.
To illustrate the ability of OKBR to conduct analysis on an inordinately large dataset, 
I also repeat the OKBR estimation for a simulated time-series with~$N=10^{10}$ data-points. 
For all three cases, I estimate drift and diffusion coefficients from the conditional moments using direct estimation~\cite{siegert1998analysis}

\begin{equation}\label{eq:direct-estimation}
    \mathbf{\hat{D}}^{(k)} = \frac{1}{k!\Delta t}\mathbf{\hat{M}}^{(k)}.
\end{equation}
Results are shown in Fig.~\ref{figure:OU-example}.

I find that for the~$N=10^7$ case, KBR and OKBR give identical estimates for the drift and diffusion coefficients, and the coefficients~$|x|\lesssim 2$ are estimated fairly.
However, at the rarely sampled edges, either large errors are present or there are no samples available to make an estimate.
For the~$N=10^{10}$ case, OKBR accurately recovers the drift and diffusion coefficients over the entire estimation range.
It is not possible to use KBR on the~$N=10^{10}$ dataset, as the data does not fit within computer memory.

\subsubsection{Empirical performance of estimation procedures}

To empirically benchmark the time and space requirements for KBR and OKBR, I repeat the estimations in Sec.~\ref{ssec:OU-process-example}, varying the number of data-points~$N$ and leaving all other parameters unchanged.
The number of data-points considered is~$N\in(10^{4}, 10^{5}, \dots, 10^{10})$, however estimation using the largest dataset using KBR is not possible due to memory requirements.
Table~\ref{tab:benchmark-results} shows the benchmark results.
Both methods show linear scaling in time.
KBR shows linear scaling in space, whereas the space requirements of OKBR scale to a constant, independent of~$N$.

\begin{table}[b]
\caption{\label{tab:benchmark-results}%
Time and space requirements for KBR and OKBR, varying the number of data points~$N$ in the time-series dataset. 
Estimation using KBR is not possible for the~$N=10^{10}$ dataset due to memory restrictions.
Similar scalings were found for examples~\ref{ssec:tri-stable-example} and~\ref{ssec:mult-corr-example}. For KBR, the memory estimate is dominated by the size of the time series itself.
}
\begin{ruledtabular}
\begin{tabular}{ccccc}
& \multicolumn{2}{c}{KBR} & \multicolumn{2}{c}{OKBR} \\
\cmidrule{2-3} \cmidrule{4-5}
\textrm{N} & \textrm{Time (s)} & \textrm{Space (GB)} & \textrm{Time (s)} & \textrm{Space (GB)}\\
\colrule
${10}^4$    &~$4.49\times 10^{-2}$ &~$1.28\times 10^{-4}$ & 
   ~$4.08\times 10^{-2}$ &~$7.44\times 10^{-5}$ \\
${10}^5$    &~$3.89\times 10^{-1}$ &~$8.48\times 10^{-4}$ & 
   ~$3.65\times 10^{-1}$ &~$1.03\times 10^{-4}$ \\
${10}^6$    &~$4.05\times 10^{0}$ &~$8.05\times 10^{-3}$ & 
   ~$3.65\times 10^{0}$ &~$1.63\times 10^{-4}$ \\
${10}^7$    &~$4.05\times 10^{1}$ &~$8.00\times 10^{-2}$ & 
   ~$3.64\times 10^{1}$ &~$1.63\times 10^{-4}$ \\
${10}^8$    &~$4.08\times 10^{2}$ &~$8.00\times 10^{-1}$ & 
   ~$3.64\times 10^{2}$ &~$1.64\times 10^{-4}$ \\
${10}^9$    &~$4.06\times 10^{3}$ &~$8.00\times 10^{0}$ & 
   ~$3.64\times 10^{3}$ &~$1.64\times 10^{-4}$ \\
${10}^{10}$    &  &  & 
   ~$3.64\times 10^{4}$ &~$1.64\times 10^{-4}$ \\
\end{tabular}
\end{ruledtabular}
\end{table}

\subsection{\label{ssec:tri-stable-example}Tri stable system}

I consider a system which exhibits poorly sampled regions of state space, arising from fast, transient dynamics through unstable (or metastable) states.
One natural example of such a system is the time-variability of the axial dipole moment of Earth's geomagnetic field, which shows two prominently stable states at positive and negative polarity, and an unstable (or possibly metastable) ``weak'' state during polarity transitions~\cite{constable1988statistics,lhuillier2013statistical,wicht2016gaussian}.
The qualitative dynamics of this system can be represented by the toy model

\protect 
\begin{subequations}\label{eq:example2-dd}
\begin{align}
    D^{(1)}(x) &= - x + 27x^3 -26x^5,\\
    D^{(2)}(x) &= \frac{7}{10}.
\end{align}
\end{subequations}

This system is characterized by two strong attractors at~$x=\pm 1$, and one weaker, rarely sampled attractor at~$x=0$.
Here, one might aim to determine the stability of the middle state from empirical data.

\begin{figure}
\noindent
\includegraphics[width=0.45\textwidth]{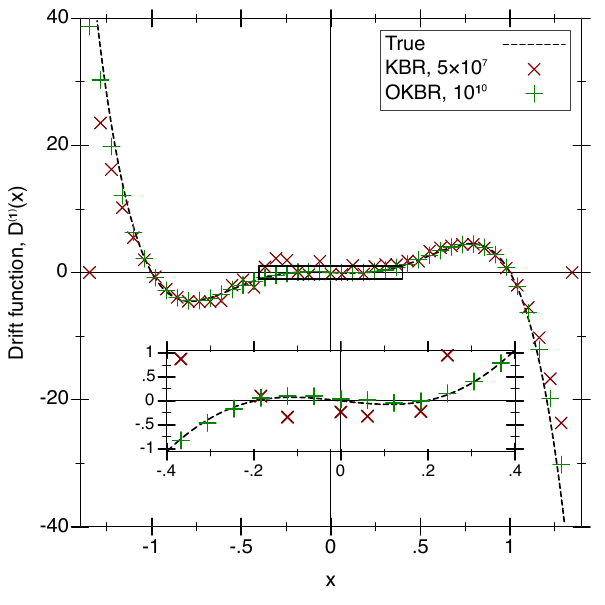}
\includegraphics[width=0.45\textwidth]{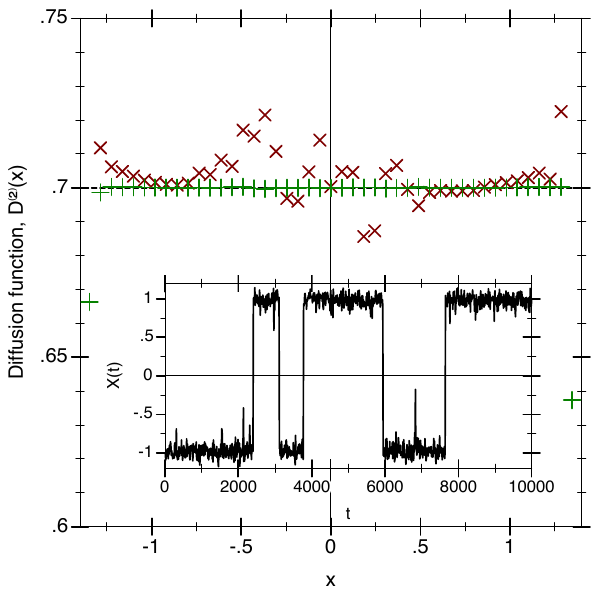}
\caption{Drift (upper) and diffusion (lower) estimates for example~\ref{ssec:tri-stable-example}. 
The inset in the upper plot shows a zoomed in section close to the origin.
The inset in the lower plot shows a section of the time-series data.
The true drift and diffusion functions are indicated with dashed black lines.
Estimates using KBR with~$N=5\times10^7$ data points are shown with red ``$\times$''~crosses.
Estimates using OKBR with~$N=10^{10}$ data-points are shown with green ``$+$''~crosses. 
A repeated estimation using OKBR and~$N=5\times10^7$ data points gives identical results to KBR, and is not plotted for conciseness.
}
\label{figure:tri-stable-example}
\end{figure}

I integrate system~\eqref{eq:example2-dd} with a sampling interval of~$\Delta t=10^{-4}$, using~$N=5\times10^7$ and~$N=10^{10}$ points for KBR~\eqref{eq:KBR} and OKBR~\eqref{eq:OKBR}, respectively.
I estimate conditional moments at 45 equally spaced points in the interval~$[-1.4,1.4]$ using a bandwidth of~$h=0.03$, and perform sampling in~$\tau$ at a series of time steps~$\bm{\mathcal{T}}=\left[\Delta t, 2\Delta t, 3\Delta t, 4\Delta t\right]^T$.
I estimate drift and diffusion coefficients in the~$\tau\rightarrow 0$ limit in~\eqref{eq:kramers-moyal-coefficients} by minimizing

\begin{equation}\label{eq:linear-regression3}
    V\left(\mathbf{\hat{D}}^{(k)}\right) = \left|\left|\mathbf{\hat{M}}^{(k)} - \bm{\mathcal{T}}\mathbf{\hat{D}}^{(k)}\right|\right|^2,
\end{equation}
using ordinary least squares.
Results are shown in Fig.~\ref{figure:tri-stable-example}.

I find that for the~$N=5\times10^7$ case, KBR is able to reasonably recover the drift and diffusion coefficients close to the attractors at~$x=\pm 1$. 
However, poor estimates are made for the rarely sampled transitions, for~$x \in [-0.5,0.5]$, and the details of stability at~$x=0$ are unresolvable. 
For the~$N=10^{10}$ case, OKBR accurately recovers the drift and diffusion coefficients across the entire sampling domain, revealing the presence of the weak attractor at~$x=0$. 

\subsection{\label{ssec:mult-corr-example}Multiplicative and correlated noise}

I consider a system with a multiplicative diffusion term and an exponentially correlated noise source~$\eta(t)$,

\begin{subequations}\label{eq:example3-process}
\begin{align}
    \frac{d}{dt}X &= D^{(1)}(X) + \sqrt{2D^{(2)}(X)}\eta(t),\\
    \frac{d}{dt}\eta &= -\frac{1}{\theta}\eta + \frac{1}{\theta}\Gamma(t),
\end{align}
\end{subequations}
where 

\begin{subequations}\label{eq:example3-dd}
\begin{align}
    D^{(1)}(x) &=  - \frac18 - \frac94 x - \frac{4}{15} x^3,\\
    D^{(2)}(x) &= 1 + \frac{1}{50}x^2 + \frac{1}{40}x^4,
\end{align}
\end{subequations}
and~$\theta=0.01$ is the correlation time of the noise~$\eta(t)$, and~$\Gamma(t)$ is internal Gaussian white noise.
Only the time-series of~$X(t)$ is observed.

\begin{figure}
\noindent
\includegraphics[width=0.45\textwidth]{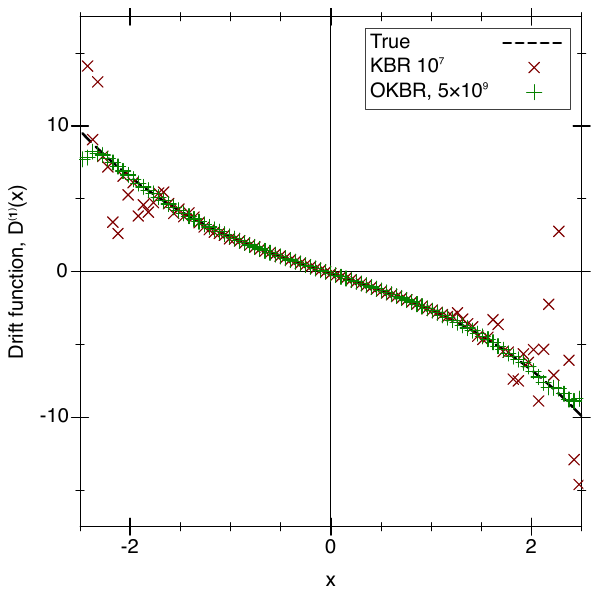}
\includegraphics[width=0.45\textwidth]{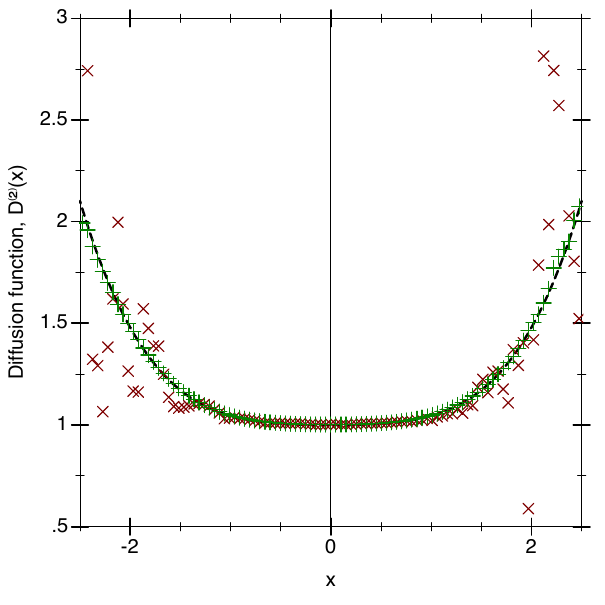}
\caption{Drift (upper) and diffusion (lower) estimates, for example~\ref{ssec:mult-corr-example}. 
The true drift and diffusion functions are indicated with dashed black lines.
Estimates using KBR with~$N=10^7$ data points are shown with red ``$\times$''~crosses.
Estimates using OKBR with~$N=5\times10^{9}$ data points are shown with green ``$+$''~crosses. 
A repeated estimation from~$N=10^7$ data points using OKBR gives identical results to KBR, and is not plotted for conciseness.
}
\label{figure:mult-corr-example}
\end{figure}

I analyze process~\eqref{eq:example3-process} and~\eqref{eq:example3-dd} using the non parametric inversion method of~\citet{lehle2018analyzing}, assuming that the timescale~$\theta$ has already been estimated~(e.g., using~\cite{day2020onlinestats}). 
This method requires estimation of the sample conditional mean---$k=1$ in~\eqref{eq:KBR} and \eqref{eq:OKBR}---as well as the conditional variance, \eqref{eq:KBR-variance} and \eqref{eq:Mstar-relation}. 
I integrate the process with a sampling interval of~$\Delta t=5\times10^{-3}$, using~$N=10^7$ and~$N=5\times10^{9}$ points for KBR and OKBR, respectively.
I estimate the conditional quantities~$\mathbf{\hat{M}}^{(k)}$ at 100 equally spaced points in the interval~$[-2.5,2.5]$ using a bandwidth of~$h=0.01$, and perform sampling in~$\tau$ using 25 time steps,~$\bm{\mathcal{T}}=\left[\Delta t, \dots, 25\Delta t\right]^T$.
To estimate the drift and diffusion coefficients using the method of~\citet{lehle2018analyzing},
I decompose the sample conditional mean and variance into basis functions~$r_i(\tau, \theta)$ and coefficients~$\lambda_i^{(k)}(x)$, given by 

\begin{equation}\label{eq:corr-M-est}
    M^{(k)}(x, \tau) \approx \sum_{i=1}^3 \lambda_i^{(k)}(x) r_i(\tau, \theta).
\end{equation}
Here, the basis functions are 

\begin{equation}\label{eq:rBasisFunctions}
    \begin{aligned}
        r_1(\tau; \theta) =& \tau - \theta(1-e^{-\tau/\theta}),\\
        r_2(\tau; \theta) =& \tau^2/2 - \theta r_1(\tau; \theta),\\
        r_3(\tau; \theta) =& \tau^3/6 - \theta r_2(\tau; \theta),
    \end{aligned}
\end{equation}
and are expressed in matrix form with elements

\begin{equation}\label{eq:R-matrix}
    R_{ij}:=r_j(\mathcal{T}_i).
\end{equation}
I solve for the coefficients by minimizing

\begin{equation}\label{eq:linear-regression-corr3}
    V\left(\bm{\lambda}^{(k)}\right) = \left|\left|\mathbf{\hat{M}}^{(k)} - \bm{R}\bm{\lambda}^{(k)}\right|\right|^2,
\end{equation}
using ordinary least squares. 
Finally, I use the~$i=1$ components of the coefficients to solve differential algebraic equations for estimates of the drift and diffusion coefficients~$\mathbf{\hat{D}}^{(k)}$; see~\citet{lehle2018analyzing} for details.
The estimated drift and diffusion coefficients are shown in Fig.~\ref{figure:mult-corr-example}.

I find that for the~$N=10^7$ case, KBR is able to recover the drift and diffusion coefficients in the range~$x \in [-1,1]$, but poor estimates are made in the rarely sampled tails.
For the~$N=5\times10^{9}$ case, OKBR is able to accurately recover the drift and diffusion coefficients over a much larger range.
To illustrate the consequences of poorly resolved tails, I use~$\mathbf{\hat{D}}^{(k)}$ to estimate the parametric coefficients of the diffusion function, 

\begin{equation}\label{eq:biquartic}
    D^{(2)}(x) = A + Bx^2 + Cx^4.
\end{equation}
Parameter estimates in Table~\ref{tab:example3-fit-results} show that both the quadratic and quartic coefficients are poorly resolved for the KBR case, with uncertainty intervals overlapping zero.
However, the increased resolution that OKBR enables results in accurate parameter estimation.

\begin{table}[b]
\caption{\label{tab:example3-fit-results}%
Fit results, with~$2\sigma$ uncertainties and~$R^2$ values.}
\begin{ruledtabular}
\begin{tabular}{c|l|l|l|c}
\multicolumn{1}{c}{N} & \multicolumn{1}{c}{A} & \multicolumn{1}{c}{B} & \multicolumn{1}{c}{C} & \multicolumn{1}{c}{$R^2$}\\
\colrule
True              &~$1.000$ &~$0.020$ &~$0.025$ &  \\
${10}^7$        &~$0.946\pm 0.133$ &~$0.113\pm 0.133$ &~$0.002\pm 0.024$ &~$0.301$ \\
$5\times{10}^9$ &~$1.001\pm 0.005$ &~$0.023\pm 0.005$ &~$0.024\pm 0.001$ &~$0.998$ \\
\end{tabular}
\end{ruledtabular}
\end{table}

\section{\label{sec:application}Application to Turbulence Data}

To illustrate one possible application of OKBR, I examine a turbulence dataset from~\citet{fuchs2022open}.
This dataset---originally published by~\citet{renner2001experimental}---comes from a turbulent air jet experiment, where time-variable observations of local air velocity were made using hot-wire measurements.
The dataset comprises~$N=1.25\times 10^7$ points sampled at 8~kHz, although other turbulence datasets can be orders of magnitude larger~\cite{fuchs2017integral}. 

The data can be used to investigate a statistical description of a turbulent cascade~\cite{peinke2019fokker}.
The measurements, under the assumption of Taylor's hypothesis of frozen turbulence, reflect spatial velocity variations $u(x)$.
Increments of these velocity variations

\begin{equation}\label{eq:obs-turb}
    \xi_{n,i} := \xi(x_n,r_i) = u(x_n) - u(x_n - r_i),
\end{equation}
define a ``zooming-in'' process in~$\xi$ for decreasing~$r$.
Following the phenomenological model of~\citet{friedrich1997description}, velocity increments evolve as a Markov process in scale~$r$. 
From this, the turbulent cascade is interpreted as a stochastic process described by a Fokker-Planck equation evolving through a sequence of velocity increments~$\xi_{n,0},\xi_{n,1},\xi_{n,2},\dots,$ at increasingly smaller~$r_0>r_1>r_2>\dots$ scales.
One can use the empirical velocity measurements to not only verify the Markov property of~$\xi(r)$, but also to estimate the corresponding drift and diffusion coefficients~\cite{peinke2019fokker}.

The conditional moments for two increment scales separated by~$\delta$ are defined as
\begin{multline}\label{eq:finite-time-CM-turb}
    M^{(k)}(\delta,\xi,r,u_N) = \\
    \int_{-\infty}^\infty [\xi^\prime(r-\delta,u_N) - \xi(r,u_N)\big]^k p(\xi^\prime| \xi,u_N)\ d\xi^\prime,
\end{multline}
for~$k=1,2$. Then, the KM coefficients are given by~\cite{peinke2019fokker}

\begin{equation}\label{eq:finite-time-KM-turb}
    D^{(k)}(\xi,r,u_N) = \frac{r}{k!}\lim_{\delta\rightarrow 0}\frac{1}{\delta}M^{(k)}(\delta,\xi,r,u_N)
\end{equation}
Analogously to~\eqref{eq:OKBR}, the online formulas for the estimator of \eqref{eq:finite-time-CM-turb} can be written as

\begin{multline}\label{eq:OKBR-turb}
    \hat{M}^{(k)}_{ij}\big|_{N} = \hat{M}^{(k)}_{ij}\big|_{N-1} + K_h(\mathcal{X}_j - \xi_{N,0})\\
    \times\left(\left[\xi_{N,i} - \xi_{N,0}\right]^k - \hat{M}^{(k)}_{ij}\big|_{N-1}\right)\Big/W_{ij}\big|_{N},
\end{multline}
where

\begin{equation}\label{eq:OKBR-W-turb}
    W_{ij}\big|_{N} = W_{ij}\big|_{N-1} + K_h\left(\mathcal{X}_j - \xi_{N,0}\right).
\end{equation}

\begin{figure}
\noindent
\includegraphics[width=0.45\textwidth]{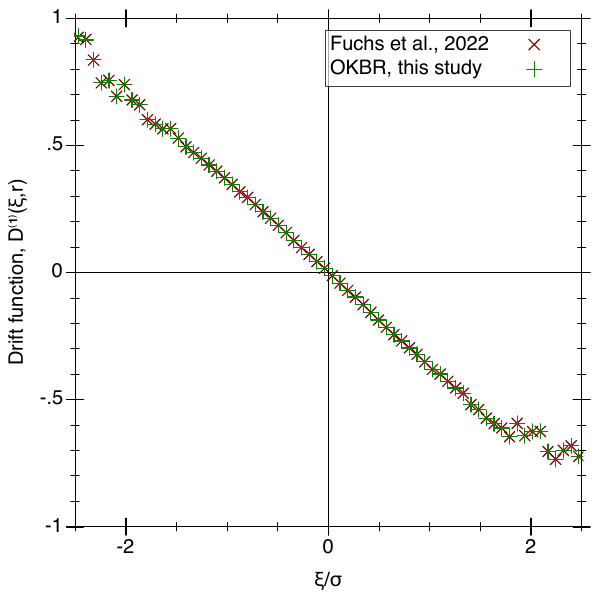}
\includegraphics[width=0.45\textwidth]{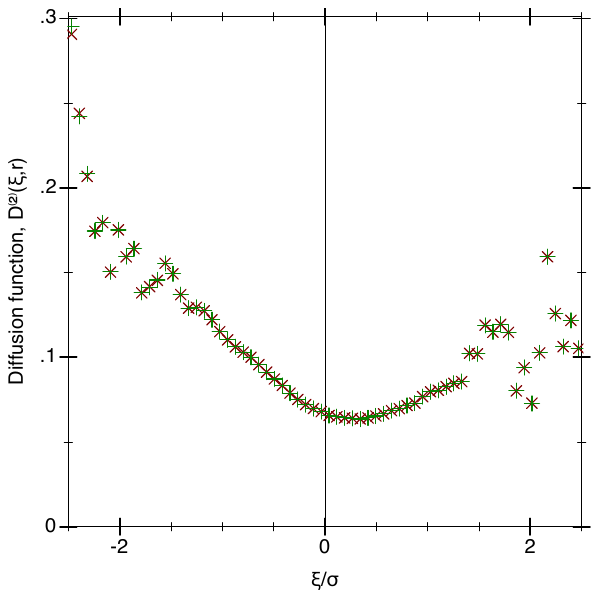}
\caption{Drift (upper) and diffusion (lower) estimates, for example~\ref{sec:application}. The coefficients $D^{(k)}(\xi,r)$ are estimated with respect to velocity increment $\xi$ for a fixed scale $r=2.7\lambda=3.2\Delta_{EM}$, where $\lambda$ is the Taylor length scale. 
Estimates from \citet{fuchs2022open} are shown with red ``$\times$''~crosses.
Estimates using OKBR are shown with green ``$+$''~crosses. 
}
\label{figure:application}
\end{figure}

I analyze the turbulence dataset comparably to~\citet{fuchs2022open} by normalizing the velocity by its variance, $\sigma$, and
estimating conditional moments using the same parameters described in~\citet{fuchs2022open}, their Fig.~23.
I use OKBR with a boxcar kernel and a bandwidth of $h=0.038$ to estimate conditional moments at a range of scales separated by $\delta$, from $\Delta_{EM}<\delta<2\Delta_{EM}$, where $\Delta_{EM}$ is the Einstein-Markov length.
KM coefficients are estimated in the $\delta\rightarrow0$ limit through linear extrapolation.
The estimated drift and diffusion coefficients are shown in Fig.~\ref{figure:application}, reproducing the previously determined results of~\citet{fuchs2022open}.

\section{\label{sec:disc-con}Discussion and Conclusion}

I present online updating formulas for estimating conditional moments and variance from time-series data.
These formulas enable the non-parametric estimation of drift and diffusion functions from arbitrarily large datasets, without requiring the entire set of input data to be available at once. 
I demonstrate this with three numerical examples.
Even for datasets that far exceed the working memory of most computers, OKBR is able to generate accurate estimates of drift and diffusion functions.
OKBR is also applied to a turbulence dataset, and the estimated drift and diffusion functions reproduce previously determined results.
Although OKBR's sequential calculations do inhibit vectorized optimizations~\cite{gorjao2019kramersmoyal}, the constant memory usage enables the analysis of exceedingly large scientific datasets.
This method could thus be incorporated into existing software packages~(e.g.,~\cite{rinn2016langevin,gorjao2019kramersmoyal,fuchs2022open,gorjao2023jumpdiff}).
Furthermore, if the updating step is faster than the data sampling interval---as is the case in example~\ref{sec:application}---OKBR may be used for real-time analysis of high-frequency streamed data.

The method presented here is demonstrated in one dimension; however, extensions to higher dimensions are straightforward.
Extensions cannot be assumed for higher-order conditional moments~($k>2$ in~$\hat{M}^{(k)}_{ij}$), as updating formulas for skewness, kurtosis, and other moments are nontrivial~\cite{pebay2008formulas}.
Further work should seek to extend the online framework to higher-order conditional moments, which would aid the analysis of jump-diffusion processes~\cite{anvari2016disentangling}.
Similar online approaches might also be used to estimate Einstein-Markov length-scales from real-time turbulence experiments~\cite{renner2001experimental,renner2006markov,stresing2011different}.

Although OKBR reduces the memory complexity to calculate conditional moments from~$\mathcal{O}(N)$ to~$\mathcal{O}(1)$, the time complexity remains at~$\mathcal{O}(N)$: competitive with traditional offline methods, as well as recent polynomial-time approaches~\cite{nikakhtar2023data}.
However, as detailed by~\citet{chan1982updating}, online formulas can sometimes be altered for calculation by multiple processing units in parallel.
It may thus be possible to estimate conditional moments in sub linear time~\cite{schubert2018numerically}.

\section*{Data Availability}

A prototype \texttt{Julia}  implementation of the estimation procedure is available at~\cite{william_davis_2023_8104832}. 
The dataset from~\citet{fuchs2022open} in Sec.~\ref{sec:application} is used under the GNU General Public License (GPL) version 3.

\begin{acknowledgments} 
I thank Matthias Morzfeld, Catherine Constable, Katherine Armstrong, and three anonymous reviewers for helpful discussions and comments which benefited this research. 
This work is supported by the Cecil H. and Ida M. Green Foundation's John W. Miles postdoctoral fellowship in theoretical and computational geophysics.
\end{acknowledgments}

\appendix

\section*{\label{app:sec:moment-derivation}Appendix: Derivation of Incremental Quantities}
\renewcommand\theequation{A\arabic{equation}}
\setcounter{equation}{0}

\subsection{\label{app:ssec:weights-Mk-derivation}Weights and conditional moments}

First I define the cumulative weights,

\begin{equation}\label{eq:app:W-definition}
    W_{ij}\big|_{N} := \sum\limits_{n=1}^{N-i} K_h\left(\mathcal{X}_j - X_{n}\right).
\end{equation}
This is rearranged to permit incremental updates

\begin{equation}\label{eq:app:W-updating}
    W_{ij}\big|_{N} = W_{ij}\big|_{N-1} + K_h\left(\mathcal{X}_j - X_{N-i}\right).
\end{equation}
Next I derive incremental formulas for conditional moments~\eqref{eq:KBR}.
Identifying the denominator of~\eqref{eq:KBR} as~\eqref{eq:app:W-definition} and rearranging gives

\protect 
\begin{equation}
    \hat{M}^{(k)}_{ij}\big|_{N} \cdot W_{ij}\big|_{N} = \sum\limits_{n=1}^{N-i} K_h(\mathcal{X}_j - X_{n})\big[X_{n+i} - X_{n}\big]^k.
\end{equation}

Separating the last term in the sum and substituting~\eqref{eq:app:W-updating} gives

\begin{multline}
    \hat{M}^{(k)}_{ij}\big|_{N} \cdot W_{ij}\big|_{N} = \\
            \hat{M}^{(k)}_{ij}\big|_{N-1} \cdot \Big(W_{ij}\big|_{N} - K_h(\mathcal{X}_j - X_{N-i})\Big)\\
        + K_h(\mathcal{X}_j - X_{N-i})\big[X_{N} - X_{N-i}\big]^k.
\end{multline}
Finally, dividing by~$W_{ij}\big|_{N}$ and rearranging gives

\begin{multline}\label{eq:app:OKBR}
    \hat{M}^{(k)}_{ij}\big|_{N} = \hat{M}^{(k)}_{ij}\big|_{N-1} + K_h(\mathcal{X}_j - X_{N-i})\\
    \times\left(\left[X_N - X_{N-i}\right]^k - \hat{M}^{(k)}_{ij}\big|_{N-1}\right)\Big/W_{ij}\big|_{N},
\end{multline}
as required by~\eqref{eq:OKBR}.

\subsection{\label{app:ssec:variance-derivation}Conditional variance}

An online calculation of the conditional variance~\eqref{eq:KBR-variance} is achieved through incremental updating of the quantity~$S_{ij}\big|_{N}$, the weighted sum of squares of differences from the current mean

\begin{equation}\label{eq:app:S-definition-offline}
    S_{ij}\big|_{N} := \sum\limits_{n=1}^{N-i} K_h(\mathcal{X}_j - X_n)
            \Big(\big[X_{n+i} - X_{n}\big] - \hat{M}^{(1)}_{ij}\big|_{N}\Big)^2.
\end{equation}
Derivation of an incremental formula for this expression uses~\eqref{eq:app:W-updating} and~\eqref{eq:app:OKBR}, and follows in a similar fashion to Sec.~A\ref{app:ssec:weights-Mk-derivation}:

\begin{widetext}
\begin{align}
    \phantom{S_{ij}\big|_{N}}
    &\begin{aligned}
        \mathllap{S_{ij}\big|_{N}} &= \Bigg[\sum\limits_{n=1}^{N-i} K_h(\mathcal{X}_j - X_n)\big[X_{n+i} - X_{n}\big]^2\Bigg] - \Big(\hat{M}^{(1)}_{ij}\big|_{N}\Big)^2 \cdot W_{ij}\big|_{N},
    \end{aligned}\\
    &\begin{aligned}
        &= S_{ij}\big|_{N-1} + K_h(\mathcal{X}_j - X_{N-i})\big[X_{N} - X_{N-i}\big]^2  + \Big(\hat{M}^{(1)}_{ij}\big|_{N-1}\Big)^2 \cdot \left(W_{ij}\big|_{N} - K_h(\mathcal{X}_j - X_{N-i})\right)
            - \Big(\hat{M}^{(1)}_{ij}\big|_{N}\Big)^2 \cdot W_{ij}\big|_{N},\\
    \end{aligned}\\
    &\begin{aligned}
        &= S_{ij}\big|_{N-1} + K_h(\mathcal{X}_j - X_{N-i})\bigg\{\big[X_{N} - X_{N-i}\big]^2
        - \Big(\hat{M}^{(1)}_{ij}\big|_{N-1}\Big)^2\bigg\} \\
        &\qquad\qquad\qquad\qquad - W_{ij}\big|_{N-1} \cdot \left(\hat{M}^{(1)}_{ij}\big|_{N} - \hat{M}^{(1)}_{ij}\big|_{N-1}\right)\left(\hat{M}^{(1)}_{ij}\big|_{N} + \hat{M}^{(1)}_{ij}\big|_{N-1}\right),\\
    \end{aligned}\\
    &\begin{aligned}
        &= S_{ij}\big|_{N-1} + K_h(\mathcal{X}_j - X_{N-i})\bigg\{\big[X_{N} - X_{N-i}\big]^2
        - \Big(\hat{M}^{(1)}_{ij}\big|_{N-1}\Big)^2 \\
        &\qquad\qquad\qquad\qquad - \Big(\big[X_{N} - X_{N-i}\big] - \hat{M}^{(1)}_{ij}\big|_{N-1}\Big)
        \Big(\hat{M}^{(1)}_{ij}\big|_{N} + \hat{M}^{(1)}_{ij}\big|_{N-1}\Big)
        \bigg\},\\
    \end{aligned}\\
    &\begin{aligned}
        &= S_{ij}\big|_{N-1} + K_h(\mathcal{X}_j - X_{N-i})
        \left(\left(X_N - X_{N-i}\right) - \hat{M}^{(1)}_{ij}\big|_{N-1}\right)
        \left(\left(X_N - X_{N-i}\right) - \hat{M}^{(1)}_{ij}\big|_{N}\right).
    \end{aligned}
\end{align}
\end{widetext}


%

\end{document}